\documentclass[epj]{svjour}
\usepackage{cite}

\usepackage{hyperref}
\usepackage[pdftex]{color}

\usepackage[latin1]{inputenc}       

\usepackage{graphicx}
\usepackage{amsmath}
\usepackage{amsfonts}

\usepackage{bbm}
\usepackage{ae}

\usepackage{subfigure}

\newcommand{\bra}[1]{\langle #1|}
\newcommand{\ket}[1]{|#1\rangle}

\newcommand{\mean}[1]{\langle #1 \rangle}

\newcommand{\boldgreek}[1]{\ensuremath{\mbox{\boldmath$#1$}}}

\begin{document}

\title{The Role of Power-Law Correlated Disorder in the Anderson Metal-Insulator 				Transition}
\titlerunning{Role of Power-Law Correlated Disorder in the Anderson MIT}

\author{Alexander Croy\thanks{Current address: Department of Applied Physics, Chalmers University of Technology, S-412 96 G\"oteborg, Sweden}, 
		  Philipp Cain and Michael Schreiber}
\authorrunning{A.\ Croy\and P.\ Cain\and M.\ Schreiber}

\institute{Institute of Physics, Chemnitz University of Technology, D-09107 Chemnitz, Germany}
\mail{alexander.croy@physik.tu-chemnitz.de}
\abstract{We study the influence of scale-free correlated disorder on the 
	metal-insulator transition in the Anderson model of localization. 
	We use standard transfer matrix 
	calculations and perform finite-size scaling of the largest inverse Lyapunov exponent
	to obtain the localization length for respective 3D tight-binding systems.
	The density of states is obtained from the full spectrum of eigenenergies 
	of the Anderson Hamiltonian.
	We discuss the phase diagram of the metal-insulator transition and 
	the influence of the
	correlated disorder on the critical exponents.
} 
\maketitle

%
%
\section{Introduction}
The possibility of having phase transitions in disordered systems, which contain randomness
as a central ingredient, has attracted a lot of interest over past decades.
Prototypical examples of classical and quantum disordered systems are the percolation
problem \cite{BroH57} and the Anderson model of localization \cite{And58}, respectively.
Over the years, the respective transitions -- percolation and Anderson metal-insulator transition (MIT) -- have been intensively studied \cite{Ess80,StaA92,Isi92,LeeR85,KraM93}. 
Typically, the random numbers, which represent the disorder, are taken to be uncorrelated.
However, in realistic systems, where, for example, the disorder is induced
by a complex environment surrounding the system sites, one expects to find correlations between
the random numbers. Spatial correlations can be characterized according to their
behavior on different length scales. For example, they might become irrelevant if the length scales associated with the phase transition are
larger than a characteristic correlation length. On the other hand, there is also scale-free disorder, which is found in many physical systems \cite{Isi92,PenBGH92,VidMNM96}. Here, the correlations are taking effect on all length scales.
These long-range correlations are characterized by a power-law behavior, $C(\bf{r} - \bf{r'}) \propto |\bf{r} - \bf{r'}|^{-\alpha}$. Here, $C(\bf{r} - \bf{r'})$ denotes the correlation function and $\alpha$ is the correlation exponent.

For the (classical) percolation problem it was found that the presence of scale-free
disorder has a profound influence on the percolation transition. For this situation, the extended Harris criterion \cite{Har74, Har83, WeiH83,Wei84} predicts a crossover of the critical exponent $\nu$ from its value $\nu_0$ 
for uncorrelated random numbers to $2/\alpha$ provided the decay of the correlations is sufficiently weak, $\alpha < 2/\nu_0$.

In the present paper we address the question of the role of power-law
correlations for the Anderson transition in disordered electronic systems. 
Originally, in his seminal paper Anderson showed that extended electronic states can
become spatially localized due to the presence of uncorrelated disorder \cite{And58}. As a consequence,
the system undergoes a phase transition from a conducting phase (for extended states) to an insulating phase (for localized states), which can be characterized by a critical exponent. 
In a previous study \cite{NdaRS04} it was found, that the critical exponent
is independent of the correlation exponent for a transition at fixed energy in the center of the band, while for a transition at fixed disorder strength
the critical exponent obeys the extended Harris criterion. However, the
calculations have been performed using a modified transfer-matrix method (TMM), which consists of forward and backward TMM calculations for a quasi-one-dimensional (quasi-1D) block of length
$L_0 \sim 10^3$. This artificial periodicity might have an additional influence on
the transition. To avoid this issue, we use the standard TMM \cite{KraM93} for calculating the localization length of quasi-1D systems with a length $L=4\cdot10^5$. Subsequently performing a finite-size scaling (FSS) analysis provides us with estimates of the critical points \cite{SleO99a} for different correlation exponents. The critical points are summarized
in a phase diagram showing the influence of the correlations on the phase boundary.
This is the central result of the present paper.
Additionally, we calculate the density of states (DOS) of 3D systems in the presence of
scale-free disorder.

The paper is organized as follows. In the next section, we introduce the
Anderson model of localization and briefly summarize the main properties of the associated
phase transition. Moreover, we provide an overview of the numerical methods
we use to calculate the properties of the transition. In Sec.\ \ref{sec:results} we present
our results for transitions at fixed energy and fixed disorder. 
Then the phase-diagram of the
Anderson MIT in the presence of scale-free disorder is discussed and the relation
to the DOS is investigated. Finally, in the last section we summarize
and discuss our results.

%
%
\section{Model and Numerical Methods}

\subsection{Anderson Model of Localization with Long-Range Correlated Disorder}
The Anderson model \cite{And58,KraM93} is widely used to investigate the
phenomenon of localization in disordered materials. It is based upon a tight-binding 
Hamiltonian in site representation
\begin{equation}
    \mathcal{H} = \sum_{\bf i} \varepsilon_{\bf i} \ket{{\bf i}}\bra{{\bf i}} 
    	- \sum_{\bf i\,j}\,t_{{\bf i j}}\,\ket{{\bf i}}\bra{{\bf j}}\;,
    \label{eq:AndersonHam}
\end{equation}
where $\ket{{\bf i}}$ is a localized state at lattice site ${\bf i}$. The matrix
elements $t_{{\bf i j}}$ denote hopping integrals between states at sites ${\bf i}$
and ${\bf j}$. Typically, hopping is restricted to nearest neighbors.
The on-site potentials $\varepsilon_{\bf i}$ are random numbers, chosen according to some probability distribution $P(\varepsilon)$ characterized by the mean $\mean{\varepsilon_i}$ and the correlation function $C(|\boldgreek{\ell}|) \propto \sum_{{\bf i}} \mean{\varepsilon_{\bf i} \varepsilon_{{\bf i}+\boldgreek{\ell}}}$. However, usually the site energies are taken to be statistically independent. For example, convenient choices of 
$P(\varepsilon)$ are a box distribution of width $W$ or a Gaussian white noise distribution, both with $\mean{\varepsilon_{\bf i}}=0$ and $C(|\boldgreek{\ell}|)=\frac{W^2}{12} \delta_{|\boldgreek{\ell}|,0}$.
Other distributions have also been considered \cite{KraM93,OhtSK99,RomS03}. 
  
  For uncorrelated potentials the resulting situation may be summarized as follows 
  \cite{KraM93}: for
  strong enough disorder, $W > W_{\rm c}(0)$, all states are exponentially localized
  to a region of finite size. The extent of this region is characterized by the
  so-called localization length $\lambda$. The value of the critical disorder 
  strength $W_{\rm c}$ depends on the distribution $P(\varepsilon)$ and the
  dimension $d$ of the system. The value of 
  $W_{\rm c}$ additionally depends on the Fermi energy $E$ and the curve $W_{\rm c}(E)$
  separates localized states, $W>W_{\rm c}(E)$, from extended states, $W<W_{\rm c}(E)$, 
  in the phase diagram. If instead of $E$ the disorder strength is fixed, there will be
  a critical energy $E_{\rm c}(W)$ and states with $|E|<E_{\rm c}$ are extended and 
  those with $|E|>E_{\rm c}$ localized. The transition from extended
  to localized wave-functions at the critical point is called 
  {\it disorder driven} or Anderson MIT. 
  In the vicinity of the critical point the localization length behaves as
  \begin{equation}\label{eq:Lambda}
  	\lambda(\tau) \propto \left| \tau_{\rm c} - \tau \right|^{-\nu}\;,
  \end{equation}
  where $\tau$ is either $E$ or $W$. The critical exponent $\nu$ characterizes
  the phase transition and is expected to be universal.   
  
  In the present work we are interested in the influence of long-range
  correlated disorder potentials on the Anderson MIT. In particular,
  we study the dependence of the critical points and the critical exponents
  on the strength of the correlations.
  To this end we use random potentials generated from a Gaussian probability distribution and with a correlation function of the form
	\begin{equation}\label{eq:PLCorrAsy}
		C(\boldgreek{\ell}) \equiv \mean{\varepsilon_{\bf i} \varepsilon_{{\bf i}+\boldgreek{\ell}}} \propto |\boldgreek{\ell}|^{-\alpha}\;,
	\end{equation}
  where $\alpha$ is the correlation exponent which determines the strength of
  the correlations. In contrast to short-range correlations the power-law behavior
  in Eq.\ \eqref{eq:PLCorrAsy} does not introduce a characteristic length scale
  and therefore the disorder is said to be scale-free.
  
  In general, it is extremely complicated to obtain analytical results of
  transport properties for the Anderson model of localization. For example, only
  in the case of $d=1$ rigorous proofs of strong localization for all energies and disorder
  strengths have been given \cite{GolMP77}. Moreover, the explicit energy and 
  disorder strength dependence of the localization length for weak disorder has been derived
  \cite{Tho79,PasF92}. There are also some results for 1D systems
  with long-range correlated disorder. For energies close to the band center a
  weak disorder expansion for $\lambda$ has been derived in Ref.~\cite{IzrK99},
  which shows the dependence of the localization length on the correlations via a
  Fourier transform of the correlation function. Here, the localization length
  is proportional to $W^{-2}$.
  On the other hand at the unperturbed band edge ($|E|=2$) it was found
  \cite{DerG84,RusHW98} that $\lambda( E=2, W) \propto W^{-y}$
  with $y = 2/3$ for $\alpha=\infty$ and $y=2/(4-\alpha)$ for $\alpha \le 1$.

\subsection{Numerical Methods}\label{sec:Methods}
In order to investigate the influence of scale-free correlations on
the Anderson MIT, we first generate the correlated on-site potential
for systems of size $M\times M\times L$ using a modified Fourier
filtering method ({FFM}) \cite{MakHSS96} with one additional step.
Namely, after performing the usual FFM we shift and scale the obtained sequence
of correlated random numbers such that the mean vanishes and the
variance is $W^2/12$. 

As mentioned in the introduction, the
localization length $\lambda$ is calculated using a standard
TMM \cite{KraM93}. Thereby we use a new seed
for each parameter combination ($E$, $W$, $\alpha$, $M$). Lastly, the critical
exponent, mobility edge and critical disorder are obtained from the
FSS analysis \cite{SleO99a} based on a
higher-order expansion of Eq.\ (\ref{eq:Lambda}) at the transition ($\tau=\tau_{\rm c}$).
This procedure is outlined in appendix \ref{sec:fsscaling}.
The error resulting from the associated fitting procedure should not be seen as an 
upper (or lower) bound for the critical
parameters, but rather as a qualitative measure for the stability of
the fitting procedure. A reliable estimate of the numerical
uncertainty requires a more sophisticated error analysis
\cite{MilRS00}. 

Another issue connected with the FSS method are the
corrections to scaling. Generally, for small systems one would always
expect to find finite-size corrections and consequently one should include
corrections to scaling in the analysis. However, this also increases
the number of parameters to be fitted tremendously and makes it sometimes
complicated to find a reasonable fit. One possibility to partly
circumvent this problem is ignoring the small system sizes (in our analysis below
this means $M=5,7$)
and doing the FSS without corrections.

The DOS is obtained from the full spectrum of eigenenergies of
the 3D Anderson Hamiltonian for systems of size
$M^3$. The eigenenergies are calculated using standard matrix
diagonalization methods \cite{lapack}. Since the
accessible maximum system size is restricted by the numerical
resources, the results are averaged over a large number of disorder
realizations to decrease statistical fluctuations. Furthermore the
symmetry of the DOS with respect to $E=0$ is utilized.

%
%
\section{Numerical Results and Discussion}\label{sec:results}

\subsection{Numerical Calculations}
\begin{figure}[tb!]
		\center
		\includegraphics[width=\columnwidth]{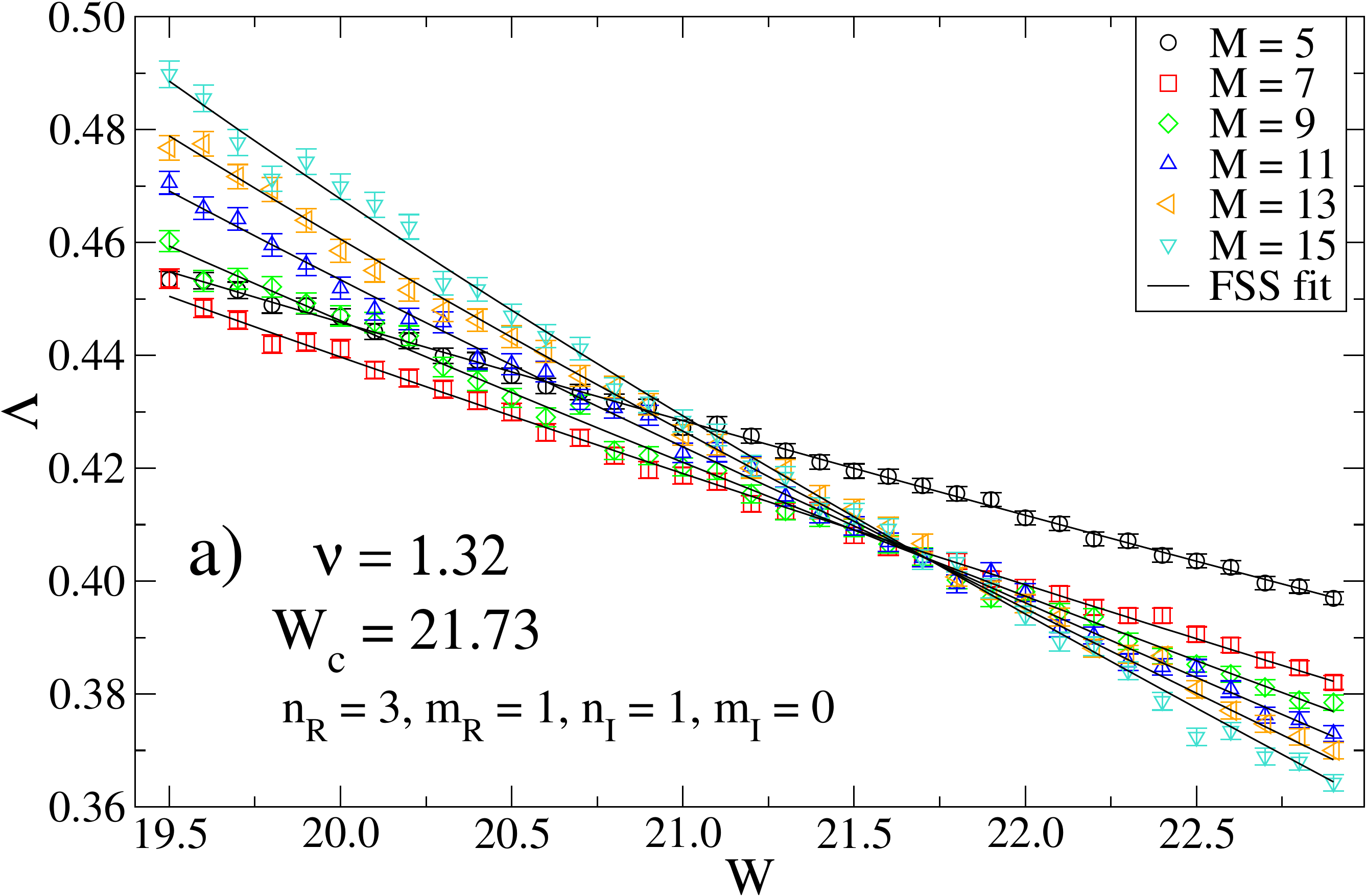}
		\includegraphics[width=\columnwidth]{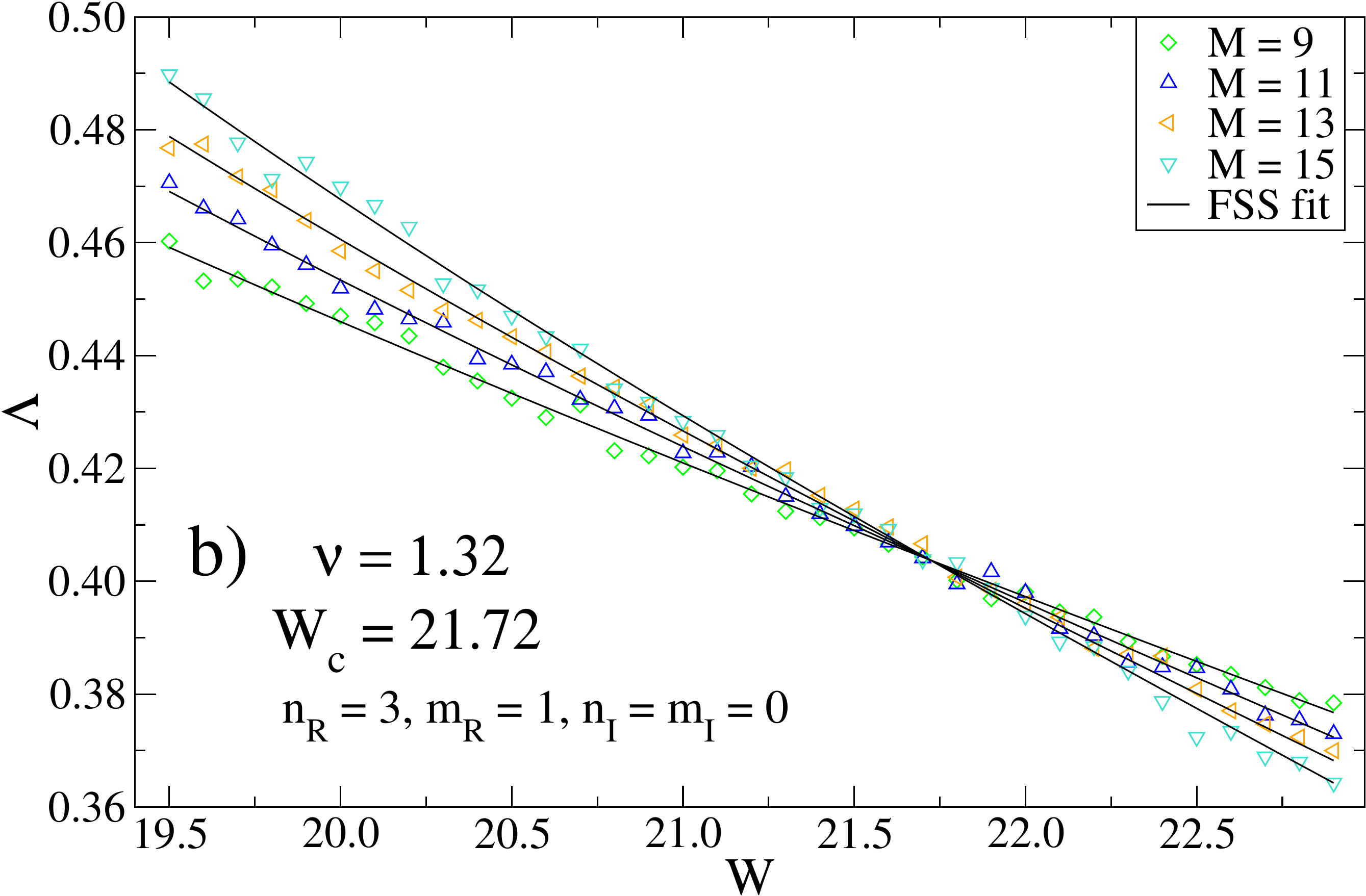}
		\caption{Reduced localization length $\Lambda$ vs disorder strength $W$
   		for $E=6.0$ and $\alpha=1.5$. 
		Solid lines show the FSS
   		fit to numerical data (a) for all system sizes $M$ taking corrections 
		to scaling into account and (b) for $M>7$ without corrections
   		to scaling. The resulting critical exponent $\nu$ and disorder
		strength $W_{\rm c}$ are shown together with the respective expansion orders of
		the FSS (cf.\ appendix \ref{sec:fsscaling}).
		}
		\label{fig:FSS-E60-A15-NoCorrSub}		
	\end{figure}

In order to study the localization length in the presence of 
correlated disorder we focus on
quasi-1D systems with $L = 400000$ and $M=5,7,9,11,13$ and $15$.
The error of the localization length is determined
from the variance of the change of the Lyapunov exponent during the
TMM iterations \cite{MacK81,MacK83}. The accuracy of the localization length
is therefore limited by the finite length of the considered systems.
To give an impression of the quality of the TMM results and the FSS
fitting procedure Fig.~\ref{fig:FSS-E60-A15-NoCorrSub} 
shows the reduced localization length $\Lambda=\lambda/M$
versus disorder strength $W$ for $E=6.0$ and $\alpha=2.5$. In
Fig.~\ref{fig:FSS-E60-A15-NoCorrSub}a the raw data for $\Lambda$
obtained from the TMM calculation are shown for various system sizes. Performing
the FSS procedure taking corrections to scaling into account one obtains
the curves shown in Fig.~\ref{fig:FSS-E60-A15-NoCorrSub}a. Considering no
corrections to scaling we obtain the fits shown in
Fig.~\ref{fig:FSS-E60-A15-NoCorrSub}b. In this case, ignoring the small 
systems leads to almost the same critical values. In the following we will
concentrate on results obtained without taking corrections to scaling into account.

The DOS is computed for disorder strengths $W=1.5,$ $\ldots,$ $30$ for the uncorrelated and a
long-ranged correlated potential ($\alpha=0.9$), respectively. The
size of the systems is $M^3=22^3$.  Results are averaged over at
least $1000$ disordered samples. For the ordered system, $W=0$, the
DOS is calculated by the diagonalization of a single system with
$M^3=30^3$.

\subsection{Transition at Fixed Energy}
First we focus on the Anderson MIT for fixed energy ($E=\rm{const.}$). 
The values of the respective critical parameters are shown in 
Tab.~\ref{tab:EnTransNoCorr}. One sees that for
uncorrelated disorder ($\alpha=\infty$) the obtained critical exponents, $\nu$, are
consistent with the high-precision value $\nu_0=1.58\pm0.03$ of Ref.~\cite{SleO99a}. 
For $E=0$ the critical disorder strength agrees very well with the value found previously \cite{SleO99a}, $W_{\rm c}=21.29\pm0.01$. 

In the presence of long-range correlations the critical exponent remains close to the
value for uncorrelated disorder potentials at least for energies well inside the band of the
system without disorder. For energies close to the band edge ($|E|=6$) of such systems the
critical exponent is smaller than $\nu_0$ in case of a long-range correlated potential.
However, at the same time the estimated error of the exponents becomes larger and is
more sensitive to changing the fitting parameters. 
The error for the critical disorder is relatively small independently of the value of 
$\alpha$ and one finds that the value is more stable than $\nu$ against changing the fitting parameters.

The value of $W_{\rm{c}}$ is monotonically increasing for decreasing $\alpha$.
In other words the MIT sets in for a larger disorder strength compared to the
uncorrelated case. This supports the intuitive expectation of an effective smoothening
of the random potential due to the correlations. For 1D systems and weak disorder
it has been shown that the effective disorder is given in terms of the Fourier transform
of the correlation function \cite{IzrK99}. In the center of the band this leads to
an increasing localization length for smaller correlation exponents \cite{CroCS11}.
\begin{table}[b]
	\center
	\begin{tabular}{@{}l@{}cccccccc}
	&$ \alpha $&$E      $&$  W_{\rm c}     $&$ \triangle W_{\rm c} $&$   \nu       $&$   \triangle \nu $\\
	\hline
	&$\infty	$&$0.0	$&$21.28	$&$0.04	$&$1.56	$&$0.08$\\
	&$\infty	$&$2.0	$&$20.600	$&$0.024	$&$1.54	$&$0.05$\\
	&$\infty	$&$4.0	$&$18.276	$&$0.029	$&$1.56	$&$0.06$\\
	&$\infty	$&$5.43	$&$14.81	$&$0.04	$&$1.55 	$&$0.08$\\
	\hline
	&$2.5	$&$0.0	$&$23.50	$&$0.04	$&$1.55	$&$0.08$\\
	&$2.5	$&$2.0	$&$22.92	$&$0.07	$&$1.57	$&$0.18$\\
	$\dagger\;$&$2.5	$&$4.0	$&$21.25	$&$0.07	$&$1.57	$&$0.11$\\
	$\ddagger\;$&$2.5	$&$5.43	$&$19.04	$&$0.05	$&$1.43	$&$0.09$\\
	\hline
	&$1.5	$&$0.0	$&$25.76	$&$0.05	$&$1.69	$&$0.22$\\
	&$1.5	$&$2.0	$&$25.37	$&$0.03	$&$1.60	$&$0.06$\\
	&$1.5	$&$4.0	$&$24.22	$&$0.05	$&$1.54	$&$0.09$\\
	&$1.5	$&$5.43	$&$22.63	$&$0.07	$&$1.45	$&$0.11$\\
	&$1.5	$&$6.0	$&$21.72	$&$0.04	$&$1.32	$&$0.05$\\
	\hline
	&$0.9	$&$0.0	$&$29.25	$&$0.08	$&$1.64	$&$0.27$\\
	&$0.9	$&$2.0	$&$28.99	$&$0.08	$&$1.61	$&$0.14$\\
	&$0.9	$&$4.0	$&$28.11	$&$0.08	$&$1.45	$&$0.11$\\
	&$0.9$&$5.43$	 &$26.82$	 &$0.07$	 &$1.38$ 	 &$0.08$\\
	&$0.9$&$7.0$ 	 &$24.72$	 &$0.13$ 	 &$1.23$	 &$0.12$\\
	\hline
	\end{tabular}
	\caption{Critical disorder $W_{\rm c}$ and exponent $\nu$ obtained 
	from FSS analysis without taking corrections to scaling into account. 
	The errors indicate the confidence interval of the fit. The symbols
	$\dagger$ and $\ddagger$ denote parameters coinciding in Tables \ref{tab:EnTransNoCorr}
	and \ref{tab:DisTransNoCorr}.}
	\label{tab:EnTransNoCorr}
\end{table}

\subsection{Transition at Fixed Disorder Strength}
\begin{table}[b]
	\center
	\begin{tabular}{@{}l@{}cccccl}
	&$ \alpha $&$W      $&$  E_{\rm c}     $&$ \triangle E_{\rm c} $&$   \nu       $&$   \triangle \nu $\\
	\hline
	&$\infty	$&$6.0		$&$6.52	$&$0.03	$&$1.49	$&$0.20$\\
	&$\infty	$&$12.0		$&$6.173	$&$0.009	$&$1.64	$&$0.03$\\
	&$\infty	$&$16.5		$&$4.855	$&$0.017	$&$1.58	$&$0.05$\\
	&$\infty	$&$19.0		$&$3.545	$&$0.019	$&$1.59	$&$0.04$\\
	\hline
	&$2.5	$&$6.0		$&$6.615	$&$0.015	$&$1.58	$&$0.06$\\
	&$2.5	$&$12.0		$&$6.826	$&$0.010	$&$1.801	$&$0.028$\\
	&$2.5	$&$16.5		$&$6.232	$&$0.011	$&$1.883	$&$0.029$\\
	$\ddagger\;$&$2.5	$&$19.0		$&$5.428	$&$0.013	$&$1.92	$&$0.03$\\	
	$\dagger\;$&$2.5	$&$21.38	$&$4.02	$&$0.03	$&$1.96	$&$0.05$\\
	\hline
	&$1.5     $&$6.0   $&$6.718   $&$0.028   $&$1.57    $&$0.09$\\
	&$1.5     $&$12.0  $&$7.484   $&$0.017   $&$1.93    $&$0.04$\\
	&$1.5     $&$16.5  $&$7.329   $&$0.023   $&$2.03    $&$0.07$\\
	&$1.5     $&$19.0  $&$7.056   $&$0.027   $&$2.30    $&$0.11$\\
	\hline
	&$0.9$    &$6.0$ 	&$6.52$ &$0.04$ &$1.64$ & $0.27$\\
	&$0.9$    &$12.0$ 	&$7.35$ &$0.03$ &$1.93$ & $0.08$\\
	&$0.9$    &$16.5$ 	&$7.71$ &$0.05$ &$2.03$  & $0.20$ \\
	&$0.9$    &$19.0$ 	&$7.70$ &$0.04$ &$2.19$ & $0.06$\\
	\hline
	\end{tabular}
	\caption{Critical energy $E_{\rm c}$ and exponent $\nu$ obtained 
	from FSS analysis without taking corrections to scaling into account. 
	The errors indicate the confidence interval of the fit.
	}
	\label{tab:DisTransNoCorr}
\end{table}
%
\begin{table}[b]
	\center	
	\begin{tabular}{cccccccccc}
	$ \alpha $&$W      $&$  E_{\rm c}     $&$ \triangle E_{\rm c} $&$   \nu       $&$   \triangle \nu $ & $y$ & $\triangle y$\\
	\hline
	$1.5	$&$12.0		$&$7.79	$&$0.07	$&$1.93	$&$0.05	$&$2.37	$&$0.24$\\
	$1.5	$&$16.5		$&$7.56	$&$0.05	$&$2.16	$&$0.05	$&$3.5	$&$0.4$\\
	$1.5	$&$19.0		$&$7.13	$&$0.05	$&$2.26	$&$0.09	$&$4.5	$&$0.8$\\
	\hline	
	\end{tabular}
	\caption{Critical energy $E_{\rm c}$ and exponent $\nu$ from FSS analysis with
	taking corrections to scaling into account. The errors indicate the confidence
	interval of the fit.}
	\label{tab:DisTransCorr}
\end{table}

Next, we consider the case of transitions at fixed disorder strength ($W={\rm const.}$). 
The respective results for critical energies and exponents are summarized in 
Tab.~\ref{tab:DisTransNoCorr}. For comparison Tab.~\ref{tab:DisTransCorr}
contains some FSS results obtained with corrections to scaling taken into
account. 

For uncorrelated random potentials previous studies showed that the critical
exponent of the transition at fixed disorder is, within error bars, identical to the exponent obtained for the transition at fixed energy.
Here, we also find a good agreement with the respective exponent shown in
Tab.\ \ref{tab:EnTransNoCorr} and with the high-precision value
$\nu_0$ \cite{SleO99a}. Further, the critical disorder strengths are in accordance
with the results of Ref.\ \cite{BulSK87}.

In contrast to the transition at fixed energy discussed earlier, in the presence
of long-range correlations the critical exponents are larger than the respective exponent 
obtained for uncorrelated potentials for all disorder strengths except $W=6.0$.
Moreover, the mobility edge is systematically shifted towards higher energies.

\subsection{Phase Diagram and DOS}
\begin{figure}[t!]
	\center
	\includegraphics[width=.95\columnwidth]{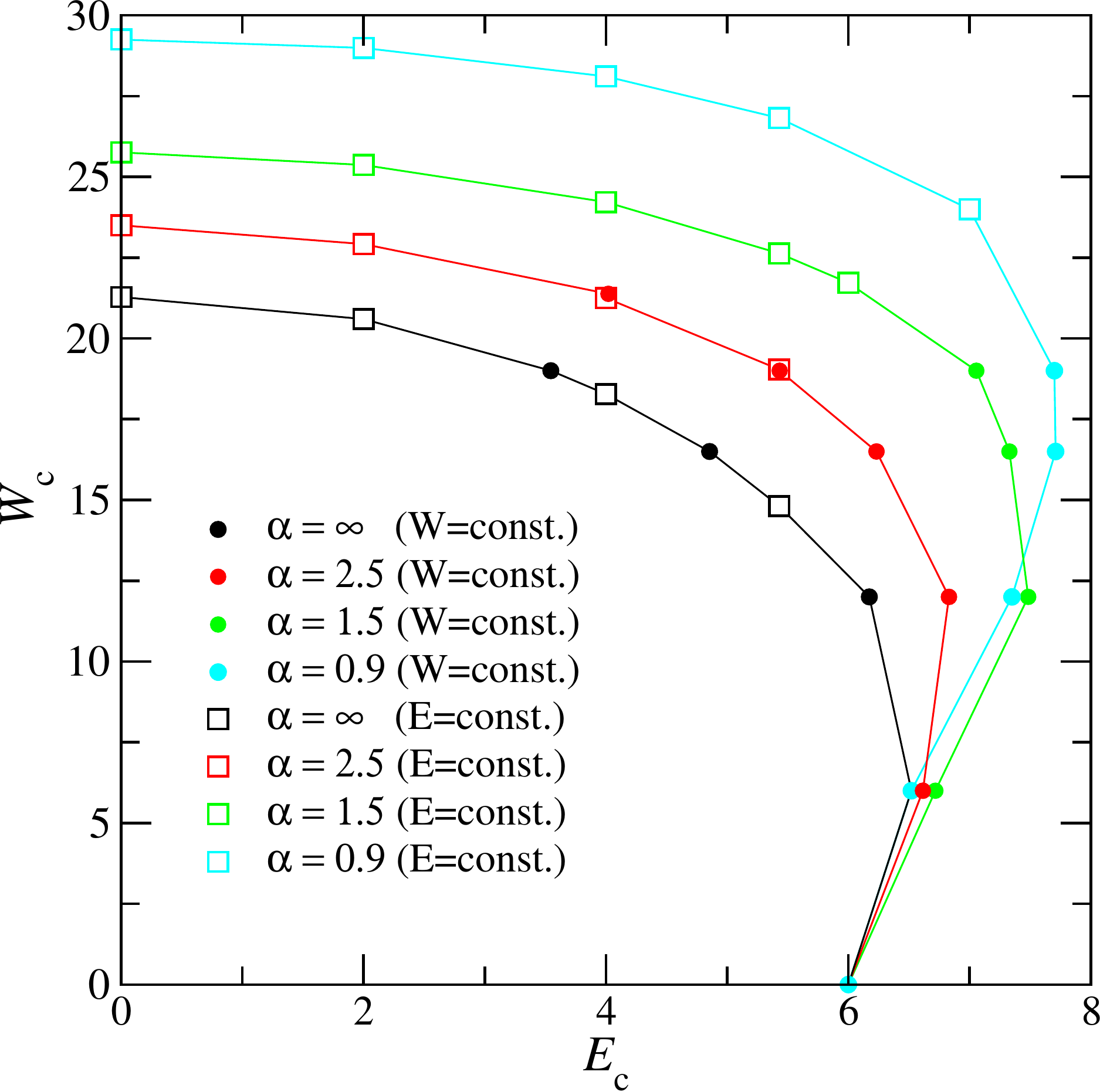}
	\caption{Phase diagram of the Anderson model with long-range correlated
	disorder. Open and filled symbols indicate
	transitions at fixed energy and fixed disorder strength, respectively. 
	Lines are a	guide to the eye only.}
	\label{fig:PhaseDiagram}
\end{figure}

Combining Tabs.~\ref{tab:EnTransNoCorr} and \ref{tab:DisTransNoCorr} we obtain a
complete phase diagram of the Anderson model in presence of long-range correlated disorder, which is shown in Fig.~\ref{fig:PhaseDiagram}. The phase diagram reflects the 
general features we have discussed for the two transitions. In the presence of
long-range correlations the metallic phase space grows, pushing the mobility edge
to larger disorder strengths and higher energies.

\begin{figure}[b!]
	\center
	\includegraphics[width=.95\columnwidth]{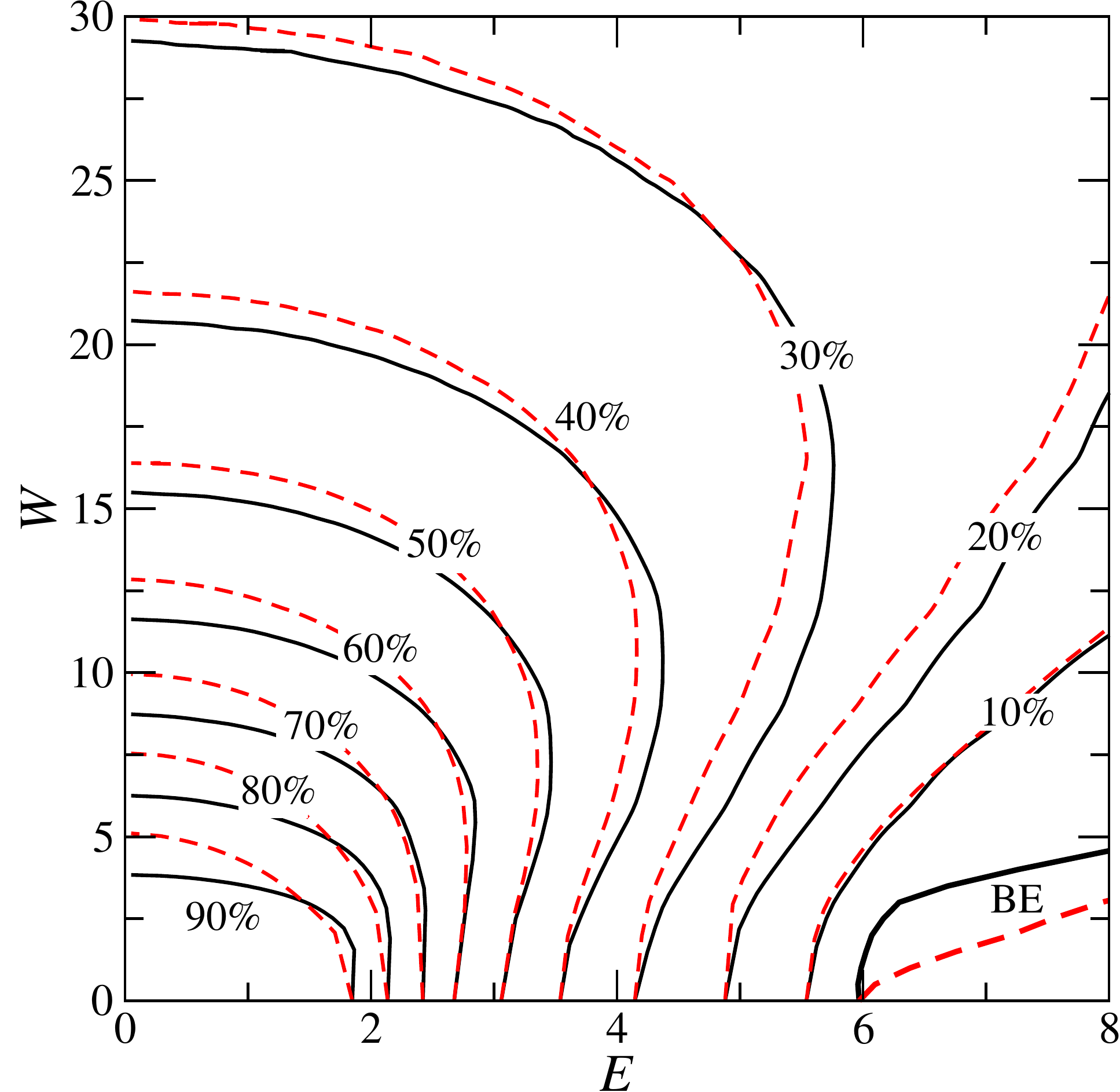}
	\caption{The contour diagram shows the disorder dependence of
          the DOS for the uncorrelated (black) and a correlated
          potential with $\alpha=0.9 $ (red dashed).  Lines are drawn at the
          given percentage of the maximum value of the DOS located at $W=0$
          and $E=0$. The band edge (BE) is estimated by the largest eigenenergy observed.  }
	\label{fig:DOS_contour}
\end{figure}

Figure \ref{fig:DOS_contour} illustrates the influence of
long-range correlated disorder on the DOS.  
The contours in Fig.\ \ref{fig:DOS_contour}
show the characteristic broadening of the DOS for increasing
disorder strength $W$ \cite{BulSK87}. The difference between
correlated and uncorrelated disorder is much less pronounced than in
the phase diagram. Around the band center ($E<2$) the DOS is
increased by correlations. Toward the band edges the DOS is slightly
smaller. From Figs. \ref{fig:PhaseDiagram} and \ref{fig:DOS_contour}
we can conclude that the mobility edge is always clearly inside the
band and comes close to the band edge only for small disorder strengths $W$.

One might also ask how the two transitions behave at the same point in the phase
diagram. A previous study of the Anderson model with uncorrelated disorder
suggested that close to the band edge and for fixed disorder strength
the critical exponent may be different from $\nu_0$ \cite{KraBMS90}. However, 
strong finite-size effects in this region did not allow a conclusive answer.
A more recent study indicates that for both transitions at different points
in the phase diagram the same exponent is obtained \cite{BrnM06}.
From Fig.~\ref{fig:PhaseDiagram} we see that there are
two pairs $(W_{\rm c},E_{\rm c})$ which denote the same phase-diagram point,
respectively. The critical parameters for these points are 
marked in Tabs. \ref{tab:EnTransNoCorr} and \ref{tab:DisTransNoCorr} by $\dagger$ and $\ddagger$. In Fig.~\ref{fig:PhaseDiagram} these points are conspicuous, because
the (red) open squares and filled circles coincide.
In both cases the transitions at fixed energy yield a critical exponent in agreement 
with $\nu_0$,
while the exponents of transitions at fixed disorder strength are larger than $\nu_0$.
A similar behavior has been reported in Ref.\ \cite{NdaRS04}, where
in addition an agreement of the critical exponents of the transition at fixed 
disorder strength with the extended Harris criterion was found. Due to the limited accuracy of our
numerical data and the sensitivity of the critical exponent, we cannot give
a quantitative comparison with the Harris criterion. Nevertheless, it is interesting to
notice, that in our case the critical exponents strongly depend on the value
of $W$ and only weakly on the value of $\alpha$. Moreover, there is
a possible connection of the behavior of $\nu$ and the slope of the curve describing the
phase boundary, $W_{\rm c}(E)$ and $E_{\rm c}(W)$, respectively. For example,
for fixed disorder strength the critical exponents increase with
increasing magnitude of $dE_{\rm c}(W)/dW$. In other words, only if the chosen path through
the phase diagram is perpendicular to the phase boundary, we obtain an unchanged
critical exponent from the FSS analysis compared to the uncorrelated value $\nu_0$.
Otherwise the estimated exponent is different from $\nu_0$.
A detailed investigation of this behavior would certainly be very interesting
and might help to elucidate the role of long-range correlations for the 
Anderson MIT.
%
%
\section{Summary and Conclusions}
In summary, we have studied the role of scale-free disorder in the Anderson MIT. The correlations are characterized by a power-law with a correlation exponent. The
characteristics of the Anderson transition have been obtained from the numerically calculated behavior of the localization length in quasi-1D systems. We employed a standard
TMM computation and estimated the critical exponents and critical points using
a FSS analysis for different correlation exponents. Further, we obtained the phase
diagram for the Anderson MIT in presence of scale-free disorder.

We observe a shift of the phase boundary towards higher energies and stronger disorder,
respectively. The latter may be understood as a result of an effective smoothening
of the disorder potential in presence of correlations. A similar behavior has been observed for 1D systems, where the localization length increases for smaller correlation
exponents.

Regarding the critical exponents we cannot draw quantitative conclusions due to the 
high sensitivity of the fitted results. However, qualitatively we see strong indications
that the critical exponents behave differently for transitions at fixed energy and fixed disorder strength as it was reported before \cite{NdaRS04}. For fixed energies $|E| > 6$ the critical exponent remains consistent with the value for uncorrelated disorder, while for 
fixed disorder strengths $W$ the exponent increases for increasing $W$. Further investigations in this direction would certainly be helpful to get a better understanding
of the role of correlations for the Anderson MIT.

%
%
\appendix
\setcounter{equation}{0}
\renewcommand{\theequation}{A.\arabic{equation}}
\section{Finite-Size Scaling} \label{sec:fsscaling}
A problem one is always faced with when using numerical methods to investigate phase transitions, is the fact that for finite systems there can be no singularities induced by a transition and the divergences are always rounded off \cite{Car96}.
However, the phase transition can still be studied using FSS.
Specifically, near the MIT one expects the following one-parameter scaling law for the
reduced localization length \cite{Car96}
\begin{equation}
    \Lambda(M,\tau,b) = \mathcal{F}\left(\frac{M}{b}, \chi(\tau) b^{1/\nu}, \phi(\tau) b^{-y}\right)\;,
    \label{eq:GeneralScaling}
\end{equation}
where $b$ is the scale factor, $\chi$ is a relevant scaling variable, $\phi$ is an irrelevant scaling variable, $\nu>0$ is the critical exponent and $y>0$ is the irrelevant scaling exponent. The irrelevant scaling variable allows us to take account of {\it corrections to scaling} due to the finite size of the sample. Here, the parameter $\tau$ measures the distance from the mobility edge $E_{\rm c}$, $\tau=|E-E_{\rm c}|/E_{\rm c}$, or the distance from the critical disorder strength $W_{\rm c}$, $\tau=|W-W_{\rm c}|/W_{\rm c}$. The choice $b = M$ leads to the standard scaling form
\begin{equation}
    \Lambda(M,\tau) = F(M^{1/\nu}\chi(\tau),M^{-y}\phi(\tau))
    \label{eq:SlevinScaling}
\end{equation}
with $F$ being trivially related to $\mathcal{F}$. For $\tau$ close to zero we expand $F$ into a Taylor series up to order $n_{\rm I}$ and obtain a series of functions $F_n$ \cite{SleO99a}
\begin{equation}\label{eq:ScalingExpansion}
      \Lambda(M,\tau)= \sum\limits^{n_{\rm I}}_{n=0} \phi^n M^{-n y} F_n(\chi M^{1/\nu})\;.
\end{equation}
Each function $F_n$ is then expanded up to order $n_{\rm R}$.
Additionally, $\chi$ and $\phi$ are expanded in terms of the small parameter $\tau$ up to order $m_{\rm R}$ and $m_{\rm I}$, respectively. This procedure gives
\begin{equation}	
    \chi(\tau) = \sum\limits^{m_{\rm R}}_{n=1} b_n \tau^n, \quad \phi(\tau) 
                      = \sum\limits^{m_{\rm I}}_{n=0} c_n \tau^n\;.
    \label{eq:VariableExpansion}
\end{equation}
From Eqs.\ (\ref{eq:SlevinScaling}) and (\ref{eq:ScalingExpansion}) one can see that a finite system size leads to a systematic shift of $\Lambda$ with $M$, where the direction of the shift depends on the boundary conditions \cite{Car96}. Consequently, the curves $\Lambda(M, \tau)$ do not intersect at the critical point $\tau=0$ for different system sizes. The term $F_0$ on the other hand shows the expected behavior. Using a least squares fit of the numerical data to Eqs.\ (\ref{eq:ScalingExpansion}) and (\ref{eq:VariableExpansion}) allows us to extract the critical parameters $\nu$, $E_{\rm c}$ and $W_{\rm c}$ \cite{RomS03,SleO99a}. For the actual orders of the expansions as given in the legends of
Figs.\ \ref{fig:FSS-E60-A15-NoCorrSub}a and \ref{fig:FSS-E60-A15-NoCorrSub}b, we have to determine, respectively, $8$ and $4$ independent
combinations of the expansion coefficients.

%
%

\end{document}